\documentclass[twocolumn,preprintnumbers,superscriptaddress,amsmath,amssymb,nofootinbib]{revtex4}

\usepackage{graphicx}
\usepackage{dcolumn}
\usepackage{bm}
\usepackage{amsmath, amssymb}
\usepackage{braket}

\allowdisplaybreaks[1]

\usepackage[normalem]{ulem}  

\renewcommand\sout{\bgroup \color{red} \ULdepth=-.5ex \ULset}

\newcommand{\PsfigA}[2]{\includegraphics[angle=-90,width=#1]{#2}}
\newcommand{\Psfig}[2]{\includegraphics[width=#1]{#2}}
\newcommand{\PsfigII}[2]{\includegraphics[scale=#1]{#2}}

\newcommand{\SUN}[1]{\text{SU} ( #1 )}

\def\Schr{Schr\"{o}dinger }

\def\mev{\text{ MeV}}

\def\fm{\text{ fm}}

\begin{document}

\preprint{}

\title{\boldmath $\bar{K} \bar{D} N$ molecular state as a ``$u u d s
  \bar{c}$ pentaquark'' in a three-body calculation}

\author{Junko~Yamagata-Sekihara}
\email{yamagata@cc.kyoto-su.ac.jp}
\affiliation{Department of Physics, Kyoto Sangyo University, 436 Motoyama Kamigamo, Kita-ku, Kyoto, 603-8555, Japan}

\author{Takayasu Sekihara} 
\affiliation{Advanced Science Research Center, Japan Atomic Energy
  Agency, 2-4 Shirakata, Tokai, Ibaraki, 319-1195, Japan}

\date{\today}

\begin{abstract} 

  We predict a new three-body hadronic molecule composed of antikaon
  $\bar{K}$, anticharm meson $\bar{D}$, and nucleon $N$ with
  spin/parity $J^{P} = 1/2^{+}$ and isospin $I = 1/2$.  This state
  behaves like an explicit pentaquark state because its minimal quark
  configuration is $u u d s \bar{c}$ or $u d d s \bar{c}$.  Owing to
  the attraction between every pair of two hadrons, in particular the
  $\bar{K} \bar{D}$ attraction which dynamically generates $D_{s 0}
  (2317)^{-}$ and $\bar{K} N$ attraction which dynamically generates
  $\Lambda (1405)$, the $\bar{K} \bar{D} N$ system is bound, and its
  eigenenergy is calculated as $3244 - 17 i \mev$ in a nonrelativistic
  three-body potential model.  We discuss properties of this $\bar{K}
  \bar{D} N$ quasibound state which emerge uniquely in three-body
  dynamics.

\end{abstract}

\maketitle

\section{Introduction}

Studying strong interactions between hadrons is one of the most
important issues in hadron physics.  The best known strong interaction
is the nuclear force between nucleons ($N$s), which generates a large
number of atomic nuclei composed of protons and neutrons.  In addition
to the nuclear force, recent interest in the strong interactions
between hadrons is to explore bound states of mesons and baryons
governed by strong interactions between them, which are so-called
hadronic molecules.  A classic example of hadronic molecule candidates
is the $\Lambda (1405)$ resonance, which may be an $S$-wave quasibound
state of antikaon ($\bar{K}$) and $N$~\cite{Dalitz:1960du}.  Recently
an analysis of the lattice QCD energy levels with an
effective-field-theory model showed that the $\Lambda (1405)$ is
dominated by the bound $\bar{K} N$ component with isospin $I =
0$~\cite{Hall:2014uca}.  The $\bar{K} N$ molecular picture for the
$\Lambda (1405)$ was supported also in Refs.~\cite{Sekihara:2014kya,
  Kamiya:2015aea} in terms of the compositeness, which is defined as
the norm of a two-body wave function for hadronic
resonances~\cite{Hyodo:2011qc, Aceti:2012dd, Sekihara:2016xnq}.
Furthermore, recent experiments in high-energy colliders such as
Belle, BaBar, BESIII, and LHCb have revealed fruitful physics in the
charm- and bottom-quark sectors.  Besides the $X$, $Y$, and $Z$
resonances as exotic candidates, the $D_{s 0}(2317)^{-}$
resonance~\cite{Aubert:2003fg, Abe:2003jk} is of interest from the
viewpoint of hadronic molecules.
Because its mass is located just below the $\bar{K} \bar{D}$
threshold, it is natural to think that the $D_{s 0}(2317)^{-}$ is an
$S$-wave $\bar{K} \bar{D}$ bound state with isospin $I = 0$.  A
dominant $\bar{K} \bar{D}$ component for the $D_{s 0}(2317)^{-}$ was
implied by theoretical calculations~\cite{Gamermann:2006nm,
  Navarra:2015iea}, and was supported also by theoretical analyses of
lattice QCD data~\cite{Mohler:2013rwa, Torres:2014vna,
  Albaladejo:2018mhb} and of experimental
data~\cite{Albaladejo:2016hae}.


We can extend discussions on hadronic molecules of two-body systems to
those of three-body systems.  In this respect, three-body systems are
not only a key to understand interactions between hadrons but also a
good ground to investigate three-body dynamics.  For example,
properties of two-body bound states may disappear in three-body bound
states, or, conversely, some properties may emerge uniquely in the
three-body bound states.  Three-body forces may become significant,
and in general its form may differ from the three-nucleon force.
Furthermore, in case that a two-body interaction depends on the energy
owing to implicit channels which do not appear as explicit degrees of
freedom, it is not trivial how to treat the energy dependence of the
two-body interaction in the three-body calculations.  We can discuss
these three-body dynamics from properties of three-body hadronic
molecules by applying and extending approaches to solve few-body
problems developed for usual atomic nuclei.

An important progress for three-body hadronic molecules takes place
recently on the $\bar{K} N N$ quasibound state.  The $\bar{K} N N$
quasibound state was predicted based on the strong attraction in the
$\bar{K} N$ system in Ref.~\cite{Akaishi:2002bg}, which was followed
by more sophisticated theoretical
calculations~\cite{Shevchenko:2006xy, Shevchenko:2007ke, Ikeda:2007nz,
  Ikeda:2008ub, Ikeda:2010tk, Dote:2008in, Dote:2008hw, Wycech:2008wf,
  Bayar:2011qj, Barnea:2012qa, Dote:2014via, Dote:2017veg,
  Dote:2017wkk, Ohnishi:2017uni}.  Eventually, the J-PARC E15
experiment very recently observed a peak structure which can be a
signal of the $\bar{K} N N$ quasibound state~\cite{Sada:2016nkb,
  Sekihara:2016vyd, Ajimura:2018iyx}.  The study of the $\bar{K} N N$
quasibound state also triggered theoretical studies of similar
three-body hadronic molecules: for instance, $\bar{K} \bar{K}
N$~\cite{Kanada-Enyo:2008wsu, Shevchenko:2015oea}, $K \bar{K}
N$~\cite{Jido:2008kp, MartinezTorres:2008kh, MartinezTorres:2009cw,
  MartinezTorres:2010zv, Xie:2010ig}, $K K
\bar{K}$~\cite{Torres:2011jt}, and $K D N$~\cite{Xiao:2011rc}.

In this study we propose a new candidate of three-body hadronic
molecules, the $\bar{K} \bar{D} N$ three-body system with spin/parity
$J^{P} = 1/2^{+}$ and isospin $I = 1/2$.  This system has two kinds of
attraction which could be essential to make the $\bar{K} \bar{D} N$
bound state.  One is the $\bar{K} N ( I = 0 )$ interaction and the
other is the $\bar{K} \bar{D} ( I = 0 )$ interaction, which
dynamically generate the $\Lambda (1405)$ and $D_{s 0} (2317)$,
respectively.  On the other hand, the $\bar{D} N$ interaction is
moderate, but some models implied that the coupling to the
$\bar{D}^{\ast} N$ channel brings attraction to the $\bar{D} N$
interaction and generates an $S$-wave $\bar{D} N ( I = 0 )$ bound
state with binding energy $\sim 1 \mev$~\cite{Yasui:2009bz,
  Gamermann:2010zz}.  To clarify whether the $\bar{K} \bar{D} N$
three-body system is bound or not, in this manuscript we will solve a
nonrelativistic three-body potential model for the $\bar{K} \bar{D} N$
system and search for the bound state.  Throughout this study, we
assume isospin symmetry for the hadron masses and interactions, and
concentrate on the zero-charge $\bar{K} \bar{D} N$ system, which
exists in the $K^{-} \bar{D}^{0} p$-$\bar{K}^{0} D^{-} p$-$\bar{K}^{0}
\bar{D}^{0} n$ coupled channels.

We here mention that the $\bar{K} \bar{D} N$ three-body system with $I
= 1/2$ has the minimal quark configuration of $u u d s \bar{c}$ or $u
d d s \bar{c}$, so this bound state, if exists, is explicitly a
``pentaquark'' state.  This is in contrast to the
charmonium-pentaquark $P_{c} (4450)$~\cite{Aaij:2015tga}, in which the
charm and anticharm quarks are hidden inside the $P_{c} (4450)$ and
the minimal quark configuration is $u u d$.

This paper is organized as follows.  In Sec.~\ref{sec:2} we construct
the two-body local potentials between $\bar{K} N$, $\bar{K} \bar{D}$,
and $\bar{D} N$ for the subsystems of $\bar{K} \bar{D} N$.  By using
this two-body local potentials, we formulate the $\bar{K} \bar{D} N$
three-body problem in Sec.~\ref{sec:3}.  In Sec.~\ref{sec:4} we show
our numerical results and discuss properties of the $\bar{K} \bar{D}
N$ system.  Section~\ref{sec:5} is devoted to the summary and
concluding remarks of this study.

\section{\boldmath Two-body systems}
\label{sec:2}

\subsection{How to construct two-body local potentials}
\label{sec:2A}

First of all, we explain how to construct the two-body interactions
for the $\bar{K} N$, $\bar{K} \bar{D}$, and $\bar{D} N$ systems.
Because we are interested in the two-body interactions in $S$ wave, we
extract the $S$-wave projected interaction and then construct the
local and orbital-angular-momentum independent potentials for the
two-body systems which reproduce the two-body phenomena in $S$ wave.

In general, the $\bar{K} N$, $\bar{K} \bar{D}$, and $\bar{D} N$
channels couple to inelastic channels, but in this study we integrate
out them so that only the $\bar{K} N$, $\bar{K} \bar{D}$, and $\bar{D}
N$ channels are explicit degrees of freedom, according to the method
in Ref.~\cite{Hyodo:2007jq}.  For instance, in the $\bar{K}
\bar{D}$-$\pi \bar{D}_{s}$-$\eta \bar{D}_{s}$ coupled channels for the
$\bar{K} \bar{D}$ interaction, the $\pi \bar{D}_{s}$ and $\eta
\bar{D}_{s}$ channels are taken as inelastic and are integrated out.
We refer to the two-body interactions in which inelastic channels are
integrated out as effective interactions.

We start with a full coupled-channels interaction in isospin basis
$V_{j k}$ with the channel indices $j$ and $k$, which is calculated in
a certain model.  We project the interaction to the $S$ wave and take
the so-called on-shell factorization~\cite{Oset:1997it}, so $V_{j k}$
depends only on the two-body center-of-mass energy $\epsilon$.  This
interaction generates the full coupled-channels scattering amplitude
$T_{j k} ( \epsilon )$ as
\begin{align}
  T_{j k} ( \epsilon )
  = & V_{j k} ( \epsilon )
  + \sum _{l} V_{j l} ( \epsilon ) G_{l} ( \epsilon ) T_{l k} ( \epsilon )
  \notag \\
  = & \left \{ \left [ 1 - V ( \epsilon ) G ( \epsilon ) \right ]^{-1}
  V ( \epsilon ) \right \} _{j k}.
\end{align}
Here, $G_{j}$ is the hadron--hadron loop function 
\begin{equation}
  G_{j} ( \epsilon ) = i \int \frac{d^{4} k}{( 2 \pi )^{4}}
  \frac{1}{k^{2} - m_{j}^{2}} \frac{1}{( P - k )^{2} - M_{j}^{2}} ,
\end{equation}
where $m_{j}$ and $M_{j}$ are masses of particles in channel $j$ and
$P^{\mu} = ( \epsilon , \, \bm{0} )$.  We calculate the loop function with
the dimensional regularization, which brings a subtraction constant
corresponding to the cutoff for the loop.

Now suppose that we explicitly treat only channel $j = 1$ and
integrate out inelastic channels $j > 1$.  In this condition, we can
calculate the effective interaction $V^{\rm eff} $ as
\begin{equation}
  V^{\rm eff} ( \epsilon )
  = \left \{ \left [ 1 - V ( \epsilon ) \tilde{G} ( \epsilon ) \right ]^{-1} 
    V ( \epsilon ) \right \} _{j=1, k=1} ,
\end{equation}
with
\begin{equation}
  \tilde{G}_{j} ( \epsilon ) = 
  \begin{cases}
    0 & ( j = 1 ) , \\
    G_{j} ( \epsilon ) & ( j > 1 ) .
  \end{cases}
\end{equation}
Physically, $V^{\rm eff}$ is the sum of the bare interaction $V_{1 1}$
and terms which include resummation of loop contributions from the
inelastic channels $j > 1$ to all orders~\cite{Hyodo:2007jq}.  Then,
the effective interaction $V^{\rm eff}$ is translated into the local
two-body potential $U ( r )$ with the relative distance $r$ in the
nonrelativistic reduction:
\begin{equation}
  U ( r ; \, \epsilon )
  = \frac{g ( r )}{4 \omega _{1} ( \epsilon ) \Omega _{1} ( \epsilon )}
  V^{\rm eff} ( \epsilon ) .
\end{equation}
Here, $g ( r )$ is a form factor defined as
\begin{equation}
  g ( r ) = \frac{1}{\pi ^{3/2} b^{3}}
  e^{- r^{2} / b^{2}} ,
\end{equation}
and
\begin{equation}
  \omega _{1} ( \epsilon ) = \frac{\epsilon ^{2} + m_{1}^{2} - M_{1}^{2}}
         {2 \epsilon} ,
  \quad
  \Omega _{1} ( \epsilon ) = \frac{\epsilon ^{2} + M_{1}^{2} - m_{1}^{2}}
         {2 \epsilon} .
\end{equation}
The range parameter $b$ can be fixed independently in three systems:
$\bar{K} N$, $\bar{K} \bar{D}$, and $\bar{D} N$.  Note that the local
potential $U$ depends on the energy of the two-body system $\epsilon$
according to the integration of the implicit channels as well as
intrinsic energy dependence of the full interaction $V_{j k}$.

The above potential $U ( r ; \, \epsilon )$ is described in isospin
basis.  The translation into the potential in particle basis is
straightforward.

\subsection{\boldmath $\bar{K} N$ system}

In the $\bar{K} N$ subsystem in $\bar{K} \bar{D} N$, we consider three
channels: $K^{-} p$, $\bar{K}^{0} n$, and $\bar{K}^{0} p$.  The former
two channels couple to each other.

We employ the Kyoto $\bar{K} N$ effective potential developed in the
above manner in Ref.~\cite{Miyahara:2015bya}, which reproduces
experimental results on the $K^{-} p$ scattering phenomena based on
chiral SU(3) coupled-channels dynamics~\cite{Ikeda:2011pi,
  Ikeda:2012au}.  The range parameter is $b = 0.38 \fm$.  The Kyoto
$\bar{K} N$ potential in its original form is written in isospin basis
as $U_{\bar{K} N ( I = 0 )} ( r; \, \epsilon )$ and $U_{\bar{K} N ( I
  = 1 )} ( r; \, \epsilon )$.  The expression in particle basis is
\begin{align}
  & U_{K^{-} p \to K^{-} p} 
  = U_{\bar{K}^{0} n \to \bar{K}^{0} n} 
  = \frac{U_{\bar{K} N ( I = 0 )} + U_{\bar{K} N ( I = 1 )}}{2} ,
\end{align}
\begin{align}
  & U_{K^{-} p \to \bar{K}^{0} n} 
  = U_{\bar{K}^{0} n \to K^{-} p} 
  = \frac{U_{\bar{K} N ( I = 0 )} - U_{\bar{K} N ( I = 1 )}}{2} ,
\end{align}
\begin{align}
  & U_{\bar{K}^{0} p \to \bar{K}^{0} p} 
  = U_{\bar{K} N ( I = 1 )} ,
\end{align}
where we omitted parameters $(r ; \, \epsilon)$ for $U$.

The $\bar{K} N ( I = 0 )$ effective potential $U_{\bar{K} N ( I = 0 )}
( r ; \, \epsilon )$ generates two $\Lambda (1405)$ poles at $\epsilon
_{\rm pole} = 1424 - 26 i \mev$ and $1381 - 81 i
\mev$~\cite{Miyahara:2015bya} as $S$-wave bound-state solutions of the
\Schr equation
\begin{align}
  & \left [ m_{K} + m_{N} - \frac{\nabla ^{2}}{2 \mu _{K N}}
    + U_{\bar{K} N ( I = 0 )} ( r ; \, \epsilon _{\rm pole} ) \right ]
  \psi 
  ( r )
  \notag \\
  & = \epsilon _{\rm pole} \psi 
  ( r ) ,
\end{align}
where $m_{K}$ and $m_{N}$ are kaon and nucleon masses, respectively,
and $\mu _{K N} \equiv m_{K} m_{N} / (m_{K} + m_{N})$ is the reduced
mass of the $\bar{K} N$ system.  Among the two $\Lambda (1405)$ poles,
the higher pole at $\epsilon _{\rm pole} = 1424 - 26 i \mev$
corresponds to the $\bar{K} N$ quasibound state in chiral
dynamics~\cite{Sekihara:2014kya, Kamiya:2015aea}.  Properties of the
$\bar{K} N$ quasibound state in Kyoto $\bar{K} N$ potential was
discussed in Ref.~\cite{Miyahara:2015bya}; we here quote that the
average of the $\bar{K} N$ distance is $\sqrt{\langle r^{2} \rangle} =
1.06 - 0.57 i \fm$ with the Gamow-vector normalization
method~\cite{Miyahara:2015bya}.

\subsection{\boldmath $\bar{K} \bar{D}$ system}

In the $\bar{K} \bar{D}$ subsystem in $\bar{K} \bar{D} N$, we consider
three channels: $K^{-} \bar{D}^{0}$, $\bar{K}^{0} D^{-}$ and
$\bar{K}^{0} \bar{D}^{0}$.  The former two channels couple to each
other.

We employ a phenomenological Lagrangian constructed in
Ref.~\cite{Gamermann:2006nm}.  An important point is that in this
model the $D_{s 0} (2317)$ state is dynamically generated by the
strong attraction of the elastic $\bar{K} \bar{D} (I = 0)$
interaction.  In isospin $I = 0$, we have $\bar{K} \bar{D}$-$\eta
\bar{D}_{s}$ coupled channels with the interaction
\begin{align}
  & V_{\bar{K} \bar{D} \to \bar{K} \bar{D} ( I = 0 )} ( \epsilon )
  \notag \\
  & = - \frac{1}{3 f_{\pi} f_{D}}
  \left [ \gamma ( \bar{t} - \bar{u} ) + \epsilon ^{2} - \bar{u} 
    + m_{D}^{2} + m_{K}^{2} \right ] ,
  \label{eq:V_KDKD0}
\end{align}
\begin{align}
  V_{\bar{K} \bar{D} \to \eta \bar{D}_{s}} ( \epsilon )
  & = V_{\eta \bar{D}_{s} \to \bar{K} \bar{D}} ( \epsilon )
  \notag \\
  & = - \frac{1}{6 \sqrt{3} f_{\pi} f_{D}}
  \left [ \gamma ( \bar{u} - \bar{t} ) - ( 3 + \gamma )
    ( \epsilon ^{2} - \bar{u} ) \right .
    \notag \\
    & \left . \phantom{\bar{u} - \bar{t}}
      - m_{D}^{2} - 3 m_{K}^{2}
      + 2 m_{\pi}^{2} \right ] ,
\end{align}
\begin{align}
  & V_{\eta \bar{D}_{s} \to \eta \bar{D}_{s}} ( \epsilon )
  \notag \\
  & = - \frac{1}{9 f_{\pi} f_{D}}
  \left [ \gamma ( - \epsilon ^{2} + 2 \bar{t} - \bar{u} ) 
    + 2 m_{D}^{2} + 6 m_{K}^{2} - 4 m_{\pi}^{2} \right ] ,
\end{align}
where $f_{\pi}$ and $m_{\pi}$ ($f_{D}$ and $m_{D}$) are decay constant
and mass of pion ($D$ meson), respectively, and $\gamma$ is the
squared ratio of the masses of the light to heavy vector mesons.  In
the above expressions, $\bar{t}$ and $\bar{u}$ are the Mandelstam
variables projected to the $S$ wave in the on-shell factorization.
Owing to the $S$-wave projection, $\bar{t}$ and $\bar{u}$ are
functions only of $\epsilon$:
\begin{align}
  \bar{t} 
  = - \frac{1}{2 \epsilon ^{2}} & \left [ \epsilon ^{4} - \epsilon ^{2}
    ( m_{a}^{2} + m_{b}^{2} + m_{c}^{2} + m_{d}^{2} ) \right .
  \notag \\
  & \left . + ( m_{a}^{2} - m_{b}^{2} ) ( m_{c}^{2} - m_{d}^{2} ) \right ] ,
\end{align}
\begin{align}
  \bar{u} 
  = - \frac{1}{2 \epsilon ^{2}} & \left [ \epsilon ^{4} - \epsilon ^{2}
    ( m_{a}^{2} + m_{b}^{2} + m_{c}^{2} + m_{d}^{2} ) \right .
  \notag \\
  & \left . - ( m_{a}^{2} - m_{b}^{2} ) ( m_{c}^{2} - m_{d}^{2} ) \right ] ,
\end{align}
where $m_{a, b, c, d}$ are masses of particles in the $a b \to c d$
reaction.  Similarly, in isospin $I = 1$, we have $\bar{K}
\bar{D}$-$\pi \bar{D}_{s}$ coupled channels with the interaction
\begin{align}
  & V_{\bar{K} \bar{D} \to \bar{K} \bar{D} ( I = 1 )} ( \epsilon ) = 0,
  \label{eq:V_KDKD1}
\end{align}
\begin{align}
  & V_{\bar{K} \bar{D} \to \pi \bar{D}_{s}} ( \epsilon )
  = V_{\pi \bar{D}_{s} \to \bar{K} \bar{D}} ( \epsilon )
  \notag \\
  & = - \frac{1}{6 f_{\pi} f_{D}}
  \left [ \gamma ( \epsilon ^{2} - \bar{t} ) + \epsilon ^{2} - \bar{u} 
  - m_{D}^{2} - m_{K}^{2} \right ] ,
\end{align}
\begin{align}
  & V_{\pi \bar{D}_{s} \to \pi \bar{D}_{s}} ( \epsilon ) = 0 .
\end{align}
Note that in our model the $\bar{K} \bar{D} (I = 1)$ interacts only
through the $\pi \bar{D}_{s}$ intermediate channel.

Then, the inelastic channels $\pi \bar{D}_{s}$ and $\eta \bar{D}_{s}$
are integrated out and $\bar{K} \bar{D}$ effective potentials in
isospin basis, $U_{\bar{K} \bar{D} ( I = 0 )} (r ; \, \epsilon)$ and
$U_{\bar{K} \bar{D} ( I = 1 )} (r ; \, \epsilon)$, are calculated in
the method in Sec.~\ref{sec:2A}.  These potentials are translated into
those in particle basis as:
\begin{align}
  U_{K^{-} \bar{D}^{0} \to K^{-} \bar{D}^{0}} 
  = & U_{\bar{K}^{0} D^{-} \to \bar{K}^{0} D^{-}}
  \notag \\
  = & \frac{U_{\bar{K} \bar{D} ( I = 0 )} + U_{\bar{K} \bar{D} ( I = 1 )}}{2} ,
\end{align}
\begin{align}
  U_{K^{-} \bar{D}^{0} \to \bar{K}^{0} D^{-}} 
  = & U_{\bar{K}^{0} D^{-} \to K^{-} \bar{D}^{0}}
  \notag \\
  = & \frac{U_{\bar{K} \bar{D} ( I = 0 )} - U_{\bar{K} \bar{D} ( I = 1 )}}{2} ,
\end{align}
\begin{align}
  U_{\bar{K}^{0} \bar{D}^{0} \to \bar{K}^{0} \bar{D}^{0}} 
  = U_{\bar{K} \bar{D} ( I = 1 )} .
\end{align}

\begin{figure}[!t]
  \centering
  \PsfigA{8.6cm}{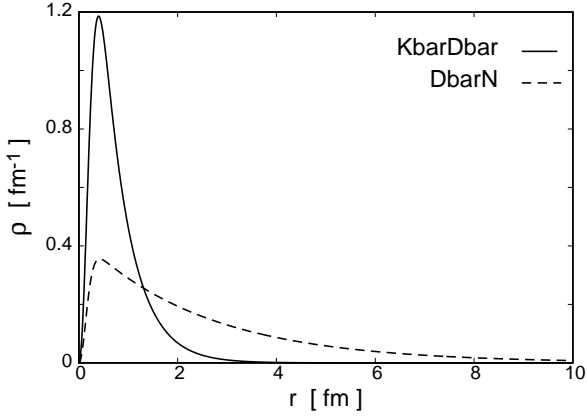}
  \caption{Density distributions for the $\bar{K} \bar{D}$ system in
    the $D_{s 0}(2317)$ (solid line) and for the $\bar{D} N$ system in
    the $\bar{D} N$ bound state (dashed line).}
  \label{fig:KD-DN}
\end{figure}

In this study, we use parameters fixed in Ref.~\cite{Navarra:2015iea}:
$\gamma = ( 800 \mev / 2050 \mev )^{2} \approx 0.152$, $f_{\pi} = 93
\mev$, and $f_{D} = 165 \mev$.  Furthermore, we fix the range of the
effective local potential as $b = 0.36 \fm$.  With these parameters,
we can generate the scalar meson $D_{s 0}(2317)^{-}$ with its mass
$\epsilon _{\rm pole} = 2317 \mev$ in the $\bar{K} \bar{D} ( I = 0 )$
channel as an $S$-wave bound-state solution of the \Schr equation with
the effective potential $U_{\bar{K} \bar{D} ( I = 0 )} ( r ; \,
\epsilon )$
\begin{align}
  & \left [ m_{K} + m_{D} - \frac{\nabla ^{2}}{2 \mu _{K D}}
    + U_{\bar{K} \bar{D} ( I = 0 )} ( r ; \, \epsilon _{\rm pole} ) \right ] \psi
  ( r )
  \notag \\
  & 
  = \epsilon _{\rm pole} \psi
  ( r ) ,
\end{align}
with the $\bar{K} \bar{D}$ reduced mass $\mu _{K D} \equiv m_{K} m_{D}
/ (m_{K} + m_{D})$.  In Fig.~\ref{fig:KD-DN} (solid line) we plot the
density distribution for the $\bar{K} \bar{D}$ system in the
$D_{s 0}(2317)$
\begin{equation}
  \rho _{\bar{K} \bar{D}} ( r )
  \equiv r^{2} \left [ \psi ( r )  \right ] ^{2} ,
  \label{eq:rho_KD}
\end{equation}
calculated from the wave function of the $\bar{K} \bar{D}$ bound state
$\psi ( r )$ with the normalization that integral of $\rho _{\bar{K}
  \bar{D}}$ with respect to $r$ in the range $[ 0, \, \infty )$ is
  unity.  The average of the $\bar{K} \bar{D}$ distance is calculated
  as $\sqrt{ \langle r^{2} \rangle _{\bar{K} \bar{D}}} = 0.93 \fm$,
  where $\langle r^{2} \rangle _{\bar{K} \bar{D}}$ is defined as
\begin{equation}
  \langle r^{2} \rangle _{\bar{K} \bar{D}}
  \equiv \int _{0}^{\infty} d r \, r^{2} \rho _{\bar{K} \bar{D}} ( r ) .
\end{equation}

\begin{figure}[!t]
  \centering
  \Psfig{8.6cm}{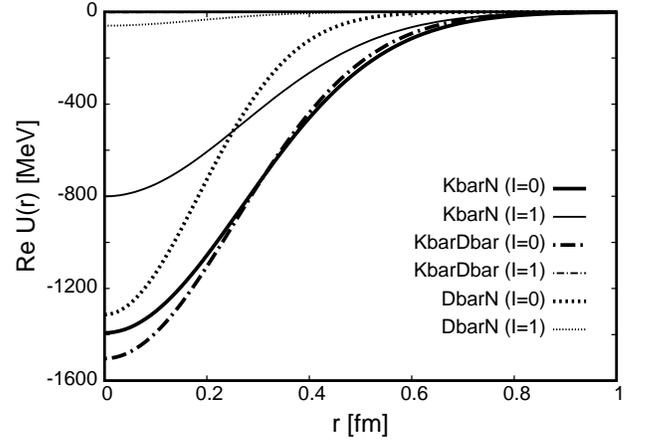}
  \caption{Real parts of the effective local potentials with the
    energies at the respective thresholds.  The $\bar{K} \bar{D} ( I =
    1 )$ potential is very tiny, as $U_{\bar{K} \bar{D} ( I = 1 )} ( r
    = 0 ) \approx - 2 \mev$.}
  \label{fig:Ueff}
\end{figure}

In Fig.~\ref{fig:Ueff}, to compare the strength of the potentials, we
show the real parts of the effective local potentials for the $\bar{K}
\bar{D} ( I = 0 )$ and the $\bar{K} \bar{D} ( I = 1 )$ systems as
thick and thin dashed-dotted lines, respectively.  The energy for the
potentials is fixed as the threshold, $\epsilon = m_{K} + m_{D}$.  As
one can see, the $\bar{K} \bar{D} ( I = 0 )$ potential has very strong
attraction, while the $\bar{K} \bar{D} ( I = 1 )$ potential is very
tiny and negligible.

We here mention that in terms of heavy quark symmetry we may have to
introduce the $\bar{D}^{\ast}$ and $\bar{D}_{s}^{\ast}$ vector mesons,
which exist $\sim 140 \mev$ above the ground $\bar{D}$ and
$\bar{D}_{s}$ mesons, respectively.  However, in this study we do not
take into account them in the $\bar{K} \bar{D}$ system because the
contributions from the $\bar{K} \bar{D}^{\ast}$, $\pi
\bar{D}_{s}^{\ast}$, and $\eta \bar{D}_{s}^{\ast}$ channels are
expected to be negligible compared to the $\bar{K} \bar{D}$ dynamics
around its threshold.

\subsection{\boldmath $\bar{D} N$ system}

In the $\bar{D} N$ subsystem in $\bar{K} \bar{D} N$, we consider three
channels: $D^{-} p$, $\bar{D}^{0} n$, and $\bar{D}^{0} p$.  The former
two channels couple to each other.

For the $\bar{D} N$ interaction, we take the approach discussed in
Ref.~\cite{Gamermann:2010zz}.  We introduce the $S$-wave channels
$\bar{D} N$ (specified by the channel $j = 1$) and $\bar{D}^{\ast} N$
($j = 2$) both in isospin $I = 0$ and $1$.\footnote{We neglect the
  $\bar{D}^{\ast} \Delta$ channel in $I = 1$, which was included in
  Ref.~\cite{Gamermann:2010zz} but was not important.}  We calculate
the interaction with a Lagrangian invariant under $\SUN{8}$ rotations
which treats heavy pseudoscalar and vector mesons on an equal footing
as required by heavy quark symmetry.  The $S$-wave interaction can be
expressed as~\cite{Gamermann:2010zz}
\begin{equation}
  V_{j k} ( \epsilon ) = \frac{\xi _{j k}}{2 f_{D}^{2}} ( \epsilon - m_{N} )
  \sqrt{[ \Omega _{j} ( \epsilon ) + m_{N} ][ \Omega _{k} ( \epsilon ) + m_{N} ]}
\end{equation}
with the on-shell nucleon energy in $j$th channel $\Omega _{j} (
\epsilon )$.  The coefficient $\xi _{j k}$ comes from the $\SUN{8}$
group structure of the couplings, whose expression is
\begin{equation}
  \xi _{(I = 0)} =
  \left (
  \begin{array}{@{\,}cc@{\,}}
    0 & - \sqrt{12} \\
    - \sqrt{12} & 4 \\
  \end{array}
  \right ) ,
  \quad
  \xi _{(I = 1)} =
  \left (
  \begin{array}{@{\,}cc@{\,}}
    2 & 4 / \sqrt{3} \\
    4 / \sqrt{3} & -2/3 \\
  \end{array}
  \right ) ,
\end{equation}
Parameters are taken from Ref.~\cite{Gamermann:2010zz}: $f_{D} = 157.4
\mev$.  An interesting feature is that, although the elastic $\bar{D}
N$ interaction is zero in $I = 0$, dynamics with the $\bar{D}^{\ast}
N$ coupled channel generates a $\bar{D} N$ bound state in $I = 0$ with
binding energy $\sim 1
\mev$~\cite{Gamermann:2010zz}.\footnote{Importance of the
  $\bar{D}^{\ast} N$ channel in the $\bar{D} N$ dynamics was pointed
  out also in Ref.~\cite{Yasui:2009bz}, where a $\bar{D} N$ bound
  state with binding energy $\sim 1 \mev$ was predicted as well.}
With such an attractive $\bar{D} N$ interaction, it is possible to
study the formation of $D$ mesic nuclei in, e.g.,
Ref.~\cite{GarciaRecio:2010vt, GarciaRecio:2011xt,
  Yamagata-Sekihara:2015ebw}.

Then we integrate out the $\bar{D}^{\ast} N$ channel and obtain
$\bar{D} N$ effective potentials in isospin basis, $U_{\bar{D} N ( I =
  0 )} (r ; \, \epsilon)$ and $U_{\bar{D} N ( I = 1 )} (r ; \,
\epsilon)$, with which we calculate the potentials in particle basis
as
\begin{align}
  & U_{D^{-} p \to D^{-} p} = U_{\bar{D}^{0} n \to \bar{D}^{0} n}
  = \frac{U_{\bar{D} N ( I = 0 )} + U_{\bar{D} N ( I = 1 )}}{2} ,
\end{align}
\begin{align}
  & U_{D^{-} p \to \bar{D}^{0} n} = U_{\bar{D}^{0} n \to D^{-} p}
  = - \frac{U_{\bar{D} N ( I = 0 )} - U_{\bar{D} N ( I = 1 )}}{2} ,
\end{align}
\begin{align}
  & U_{\bar{D}^{0} p \to \bar{D}^{0} p} = U_{\bar{D} N ( I = 1 )} .
\end{align}

As for the range parameter $b$, we fix it so as to reproduce an
$S$-wave $\bar{D} N ( I = 0 )$ bound state with $1 \mev$ binding
($\epsilon _{\rm pole} = 2805 \mev$) as a solution of the \Schr
equation with the effective potential $U_{\bar{D} N ( I = 0 )} ( r ;
\, \epsilon )$
\begin{align}
  & \left [ m_{D} + m_{N} - \frac{\nabla ^{2}}{2 \mu _{D N}}
    + U_{\bar{D} N ( I = 0 )} ( r ; \, \epsilon _{\rm pole} ) \right ] \psi
  ( r )
  \notag \\
  & 
  = \epsilon _{\rm pole} \psi
  ( r ) ,
\end{align}
with the $\bar{D} N$ reduced mass $\mu _{D N} \equiv m_{D} m_{N} /
(m_{D} + m_{N})$.  The result of the range parameter is $b = 0.26
\fm$.  In Fig.~\ref{fig:KD-DN} (dashed line) we plot the density
distribution for the $\bar{D} N$ system in the $\bar{D} N$ bound state
calculated in the same manner as in the $\bar{K} \bar{D}$ case.  The
calculated $\bar{D} N$ distance $\sqrt{ \langle r^{2} \rangle
  _{\bar{D} N}} = 3.66 \fm$ is larger than $\sqrt{ \langle r^{2}
  \rangle _{\bar{K} \bar{D}}} = 0.93 \fm$ due to the loosely bound
nature of $\bar{D} N$.

In Fig.~\ref{fig:Ueff}, we show the real parts of the effective local
potentials for the $\bar{D} N ( I = 0 )$ and the $\bar{D} N ( I = 1 )$
systems as thick and thin dotted lines, respectively.  The energy for
the potentials is fixed as the threshold, $\epsilon = m_{D} + m_{N}$.
As one can see, the $\bar{D} N ( I = 0 )$ potential is the smallest
among the $\bar{K} N$, $\bar{K} \bar{D}$, and $\bar{D} N$ potentials
with $I = 0$.  The $\bar{D} N ( I = 1 )$ potential is much smaller
than the $\bar{D} N ( I = 0 )$ potential.

\section{\boldmath $\bar{K} \bar{D} N$ three-body problem}
\label{sec:3}

\begin{figure}[!t]
  \centering
  \PsfigII{0.169}{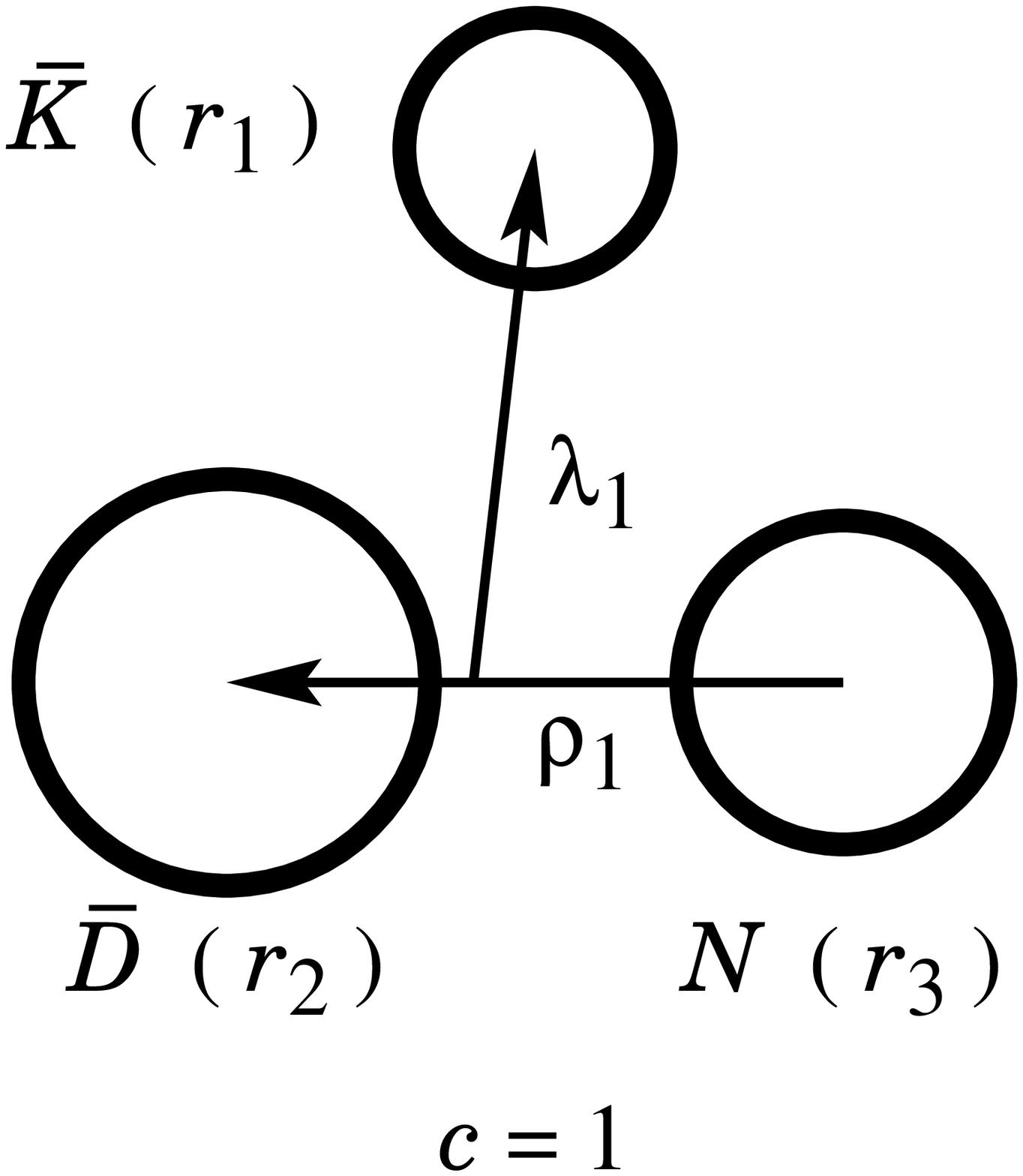}
  \PsfigII{0.169}{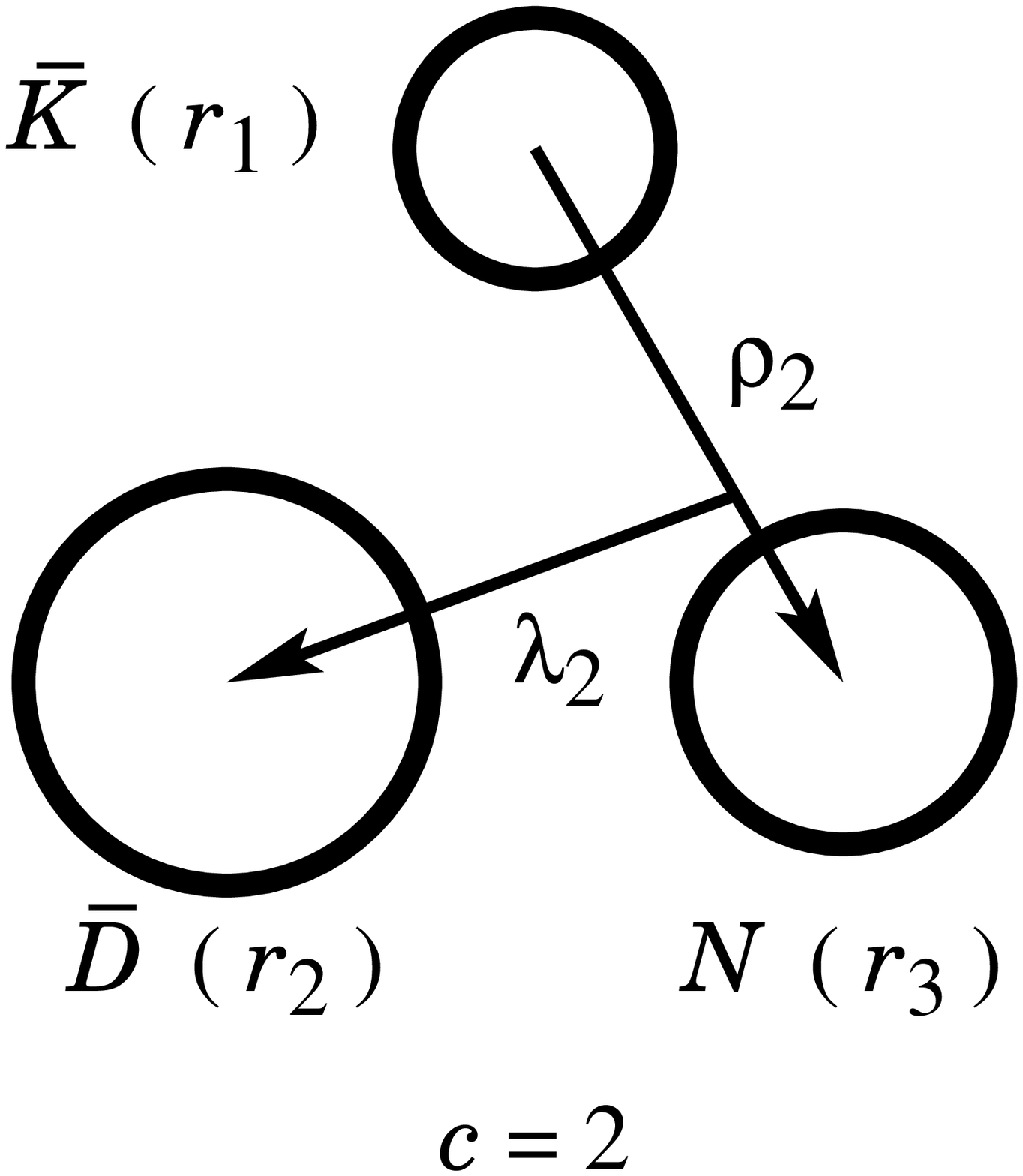}
  \PsfigII{0.169}{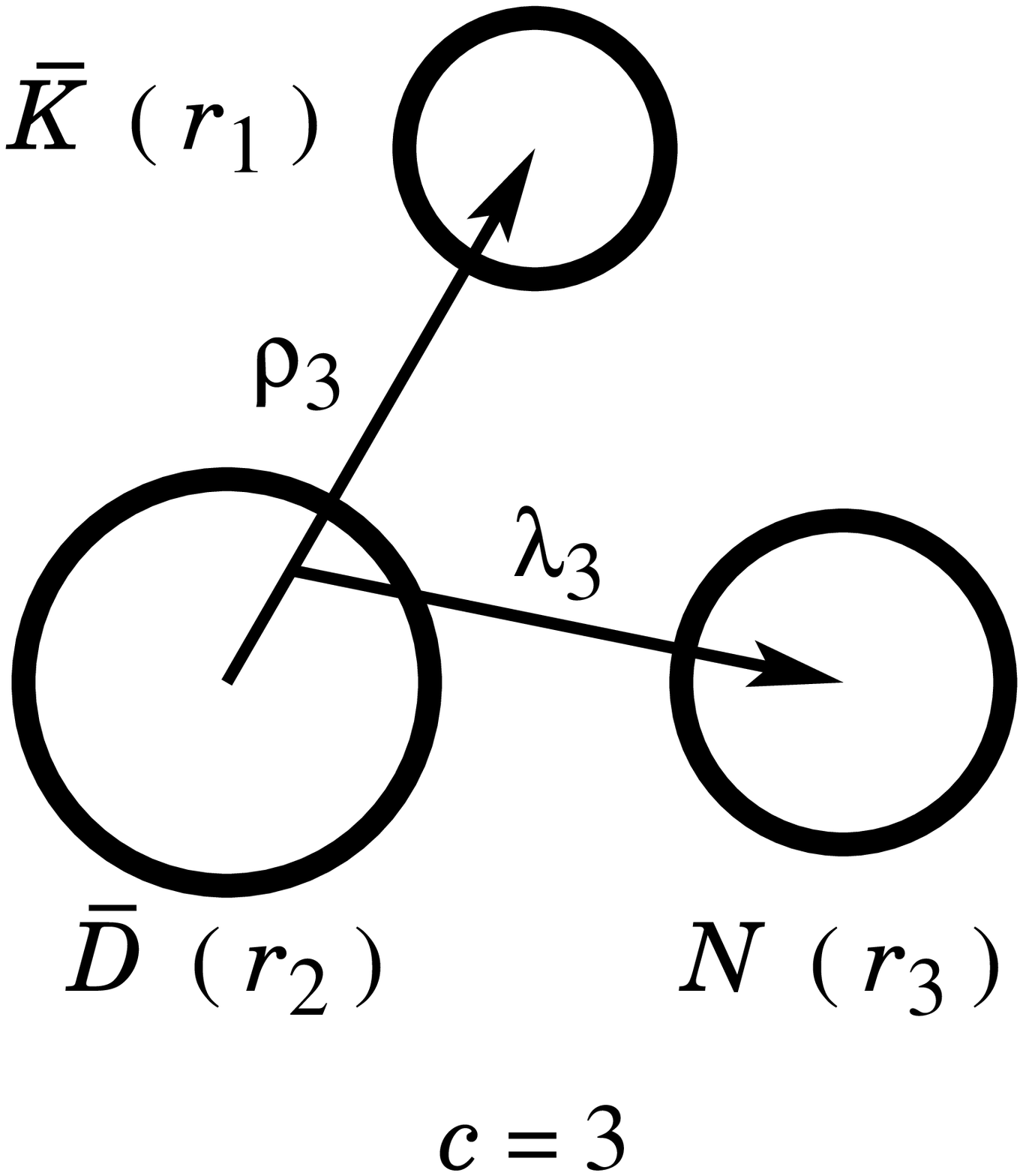}
  \caption{Three types of Jacobi coordinates.}
  \label{fig:Jacobi}
\end{figure}

Next let us formulate the $\bar{K} \bar{D} N$ three-body problem.  For
this purpose, we set the coordinates of the $\bar{K}$, $\bar{D}$, and
$N$ as $\bm{r}_{1}$, $\bm{r}_{2}$, and $\bm{r}_{3}$, respectively, and
introduce the Jacobi coordinates as in Fig.~\ref{fig:Jacobi}:
\begin{align}
  & \bm{\lambda}_{1} = \bm{r}_{1}
  - \frac{m_{D} \bm{r}_{2} + m_{N} \bm{r}_{3}}{m_{D} + m_{N}} ,
  \quad
  \bm{\rho}_{1} = \bm{r}_{2} - \bm{r}_{3} ,
  \\
  & \bm{\lambda}_{2} = \bm{r}_{2}
  - \frac{m_{N} \bm{r}_{3} + m_{K} \bm{r}_{1}}{m_{N} + m_{K}} ,
  \quad
  \bm{\rho}_{2} = \bm{r}_{3} - \bm{r}_{1} ,
  \\
  & \bm{\lambda}_{3} = \bm{r}_{3}
  - \frac{m_{K} \bm{r}_{1} + m_{D} \bm{r}_{2}}{m_{K} + m_{D}} ,
  \quad
  \bm{\rho}_{3} = \bm{r}_{1} - \bm{r}_{2} .
\end{align}
We consider the $\bar{K} \bar{D} N$ system in its center-of-mass rest
frame
\begin{equation}
  \bm{R}
  = \frac{m_{K} \bm{r}_{1} + m_{D} \bm{r}_{2} + m_{N} \bm{r}_{3}}{M_{K D N}}
  = \bm{0} ,
\end{equation}
\begin{equation}
  M_{K D N} \equiv m_{K} + m_{D} + m_{N} ,
\end{equation}
and do not treat the center-of-mass motion of the three-body system.

We employ particle basis and describe the zero-charge $\bar{K} \bar{D}
N$ system.  For the three-body system, we use index $j$ to specify the
channel in particle basis: $j = 1$ for $K^{-} \bar{D}^{0} p$, $2$ for
$\bar{K}^{0} D^{-} p$, and $3$ for $\bar{K}^{0} \bar{D}^{0} n$.  We do
not explicitly take into account other three-body channels such as
$\pi \bar{D} \Sigma$ and $\pi \bar{D}_{s} N$ but they are implemented
in the effective potentials.  We neglect the transitions to two-hadron
channels such as $\bar{D} \Lambda$ and limit our model space to the
$\bar{K} \bar{D} N$ states.

By using the Jacobi coordinates and channel index, we can express the
three-body wave function in coordinate space as
\begin{equation}
  \Psi _{j} ( \bm{\lambda}_{1} , \, \bm{\rho}_{1} ) 
  = \braket{ j ( \bm{\lambda}_{1} , \, \bm{\rho}_{1} ) | \Psi } .
\end{equation}
Here and below, the Jacobi coordinate $( \bm{\lambda}_{1} , \,
\bm{\rho}_{1} )$ is used just as a representative to specify the set
of the coordinates $( \bm{r}_{1} , \, \bm{r}_{2} , \, \bm{r}_{3} )$.

The three-body wave function $\ket{\Psi}$ satisfies the \Schr equation
\begin{equation}
  \hat{H} \ket{\Psi} = E \ket{\Psi} ,
\end{equation}
with the Hamiltonian $\hat{H}$ and an eigenvalue $E$ of the three-body
system, respectively.  Multiplying bra vector $\bra{j (
  \bm{\lambda}_{1} , \, \bm{\rho}_{1} )}$ from the left and inserting
the complete set
\begin{equation}
  1 = \sum _{k = 1}^{3} \int d^{3} \lambda _{1} \, d^{3} \rho _{1} \,
  \ket{k ( \bm{\lambda}_{1} , \, \bm{\rho}_{1} )}
  \bra{k ( \bm{\lambda}_{1} , \, \bm{\rho}_{1} )}
\end{equation}
between $\hat{H}$ and $\ket{\Psi}$ in the left-hand side, we obtain
\begin{align}
  & \sum _{k = 1}^{3} \left [ \delta _{j k} \hat{H}_{0}
    + V_{j k} ( \bm{\lambda}_{1} , \, \bm{\rho}_{1} ; \, E ) \right ]
  \Psi _{k} ( \bm{\lambda}_{1} , \, \bm{\rho}_{1} )
  \notag \\ &
  = E \Psi _{j} ( \bm{\lambda}_{1} , \, \bm{\rho}_{1} ) .
  \label{eq:Schr}
\end{align}

The kinetic term of the three-body Hamiltonian $\hat{H}_{0}$ is
\begin{align}
  \hat{H}_{0}
  = & M_{K D N} - \frac{1}{2 \mu _{1}^{\prime}}
  \left ( \frac{\partial}{\partial \bm{\lambda}_{1}} \right ) ^{2}
  - \frac{1}{2 \mu _{1}} 
  \left ( \frac{\partial}{\partial \bm{\rho}_{1}} \right ) ^{2}
\end{align}
where $\mu _{1}^{\prime}$ and $\mu _{1}$ are the reduced masses
\begin{equation}
  \mu _{1}^{\prime} = \frac{( m_{D} + m_{N} ) m_{K}}{M_{K D N}} ,
  \quad
  \mu _{1} = \frac{m_{D} m_{N}}{m_{D} + m_{N}} .
\end{equation}

As for the potential term $V_{j k}$, we employ the
orbital-angular-momentum independent potentials developed in the
previous section.  The diagonal parts $V_{j j}$ consist of all the
three combinations of two particles among $\bar{K}$, $\bar{D}$, and
$N$ in each channel:
\begin{align}
  V_{1 1} = &
  U_{\bar{D}^{0} p \to \bar{D}^{0} p}
  ( \rho _{1} ; \, \epsilon _{\bar{D} N} )
  + U_{K^{-} p \to K^{-} p} ( \rho _{2} ; \, \epsilon _{\bar{K} N} )
  \notag \\
  & + U_{K^{-} \bar{D}^{0} \to K^{-} \bar{D}^{0}}
  ( \rho _{3} ; \, \epsilon _{\bar{K} \bar{D}} ) ,
\end{align}
\begin{align}
  V_{2 2} = &
  U_{D^{-} p \to D^{-} p}
  ( \rho _{1} ; \, \epsilon _{\bar{D} N} )
  + U_{\bar{K}^{0} p \to \bar{K}^{0} p} ( \rho _{2} ; \, \epsilon _{\bar{K} N} )
  \notag \\
  & + U_{\bar{K}^{0} D^{-} \to \bar{K}^{0} D^{-}}
  ( \rho _{3} ; \, \epsilon _{\bar{K} \bar{D}} ) ,
\end{align}
\begin{align}
  V_{3 3} = &
  U_{\bar{D}^{0} n \to \bar{D}^{0} n}
  ( \rho _{1} ; \, \epsilon _{\bar{D} N} )
  + U_{\bar{K}^{0} n \to \bar{K}^{0} n} ( \rho _{2} ; \, \epsilon _{\bar{K} N} )
  \notag \\
  & + U_{\bar{K}^{0} \bar{D}^{0} \to \bar{K}^{0} \bar{D}^{0}}
  ( \rho _{3} ; \, \epsilon _{\bar{K} \bar{D}} ) ,
\end{align}
where $\epsilon _{\bar{D} N}$, $\epsilon _{\bar{K} N}$, and $\epsilon
_{\bar{K} \bar{D}}$ are energies of the subsystems $\bar{K} N$,
$\bar{D} N$, and $\bar{K} \bar{D}$, respectively, fixed later.  The
nondiagonal components of the potential consist of the charge
transition of two particles among $\bar{K}$, $\bar{D}$, and $N$:
\begin{align}
  V_{1 2} = V_{2 1} = U_{K^{-} \bar{D}^{0} \to \bar{K}^{0} D^{-}}
  ( \rho _{3} ; \, \epsilon _{\bar{K} \bar{D}} ) ,
\end{align}
\begin{align}
  V_{1 3} = V_{3 1} = U_{K^{-} p \to \bar{K}^{0} n}
  ( \rho _{2} ; \, \epsilon _{\bar{K} N} ) ,
\end{align}
\begin{align}
  V_{2 3} = V_{3 2} = U_{D^{-} p \to \bar{D}^{0} n}
  ( \rho _{1} ; \, \epsilon _{\bar{D} N} ) .
\end{align}
Because the potentials have imaginary parts according to the
implementation of the open channels, the Hamiltonian $\hat{H}$ is not
Hermitian.  Therefore, the Hamiltonian can have an eigenstate with a
complex eigenvalue, called a quasibound state.  We do not consider
three-body forces in this study.

As we have constructed in Sec.~\ref{sec:2}, the two-body potential
in the subsystem depends on its energy.  There is ambiguity to fix
energy of a two-body subsystem in a three system, but we here simply
divide the total energy $E$ among three particles according to
the ratio of masses, i.e.,
\begin{align}
  & \epsilon _{\bar{D} N} = \frac{m_{D} + m_{N}}{M_{K D N}} E ,
  \label{eq:eps_DN}
  \\
  & \epsilon _{\bar{K} N} = \frac{m_{K} + m_{N}}{M_{K D N}} E ,
  \label{eq:eps_KN}
  \\
  & \epsilon _{\bar{K} \bar{D}} = \frac{m_{K} + m_{D}}{M_{K D N}} E .
  \label{eq:eps_KD}
\end{align}
Note that the subsystem energy is complex when the total energy $E$ is
complex.

In this study we concentrate on the ground state of the $\bar{K}
\bar{D} N$ system with spin/parity $J^{P} = 1/2^{+}$, so we limit the
basis function for the three-body wave function in the channel $c$ to
having zero orbital angular momenta both for the $\lambda _{c}$ and
$\rho _{c}$ modes: $l_{\lambda _{c}} = l_{\rho _{c}} = 0$.  We then
employ the Gaussian expansion method~\cite{Hiyama:2003cu} and take the
sum of all the three rearrangements of the Jacobi coordinates, which
results in
\begin{equation}
  \Psi _{j} 
  = \sum _{c = 1}^{3} \sum _{n , n^{\prime} = 1}^{N} \mathcal{C}_{j, n n^{\prime}}^{c}
  \exp \left ( - \frac{\lambda _{c}^{2}}{r_{n}^{2}}
  - \frac{\rho _{c}^{2}}{r_{n^{\prime}}^{2}} \right ) ,
  \label{eq:WFj}
\end{equation}
with number of the expansion $N$, coefficients $\mathcal{C}_{j, n
  n^{\prime}}^{c}$, and different ranges $r_{n}$ in a geometric
progression
\begin{equation}
  r_{n} = r_{\rm min} \times
  \left ( \frac{r_{\rm max}}{r_{\rm min}} \right )^{(n - 1) / (N - 1)} .
\end{equation}
The minimal and maximal ranges, $r_{\rm min}$ and $r_{\rm max}$,
respectively, are fixed according to the physical condition of
interactions.  We comment that, although each $c$ channel in
Eq.~\eqref{eq:WFj} have zero orbital angular momentum, $l_{\lambda
  _{c}} = l_{\rho _{c}} = 0$, the sum of all the three rearrangements
allows us to take into account components with nonzero orbital angular
momenta of two-body subsystems.

By using the wave function~\eqref{eq:WFj}, the \Schr
equation~\eqref{eq:Schr} becomes
\begin{align}
  & \sum _{c = 1}^{3} \sum _{n , n^{\prime} = 1}^{N} \sum _{k = 1}^{3}
  \left \{ \delta _{j k} \left [ \frac{1}{\mu _{c}^{\prime} r_{n}^{2}}
    \left ( 3 - \frac{2 \lambda _{c}^{2}}{r_{n}^{2}} \right )
    \right . \right .
    \notag \\
    & 
    \left . \left . ~
    + \frac{1}{\mu _{c} r_{n^{\prime}}^{2}}
    \left ( 3 - \frac{2 \rho _{c}^{2}}{r_{n^{\prime}}^{2}} \right )
    + M_{K D N} - E \right ] + V_{j k} ( E ) 
  \right \}
  \notag \\
  & \times
  \mathcal{C}_{k, n n^{\prime}}^{c}
  \exp \left ( - \frac{\lambda _{c}^{2}}{r_{n}^{2}}
  - \frac{\rho _{c}^{2}}{r_{n^{\prime}}^{2}} \right ) = 0 ,
  \label{eq:GEM}
\end{align}
where we introduced the reduced masses
\begin{equation}
  \mu _{2}^{\prime} = \frac{( m_{K} + m_{N} ) m_{D}}{M_{K D N}} ,
  \quad
  \mu _{2} = \frac{m_{K} m_{N}}{m_{K} + m_{N}} ,
\end{equation}
\begin{equation}
  \mu _{3}^{\prime} = \frac{( m_{K} + m_{D} ) m_{N}}{M_{K D N}} ,
  \quad
  \mu _{3} = \frac{m_{K} m_{D}}{m_{K} + m_{D}} .
\end{equation}
Then we multiply $\exp ( - \lambda _{a}^{2} / r_{m}^{2} - \rho
_{a}^{2} / r_{m^{\prime}}^{2} )$ to the \Schr equation~\eqref{eq:GEM}
and integrate it with respect to $\bm{\lambda}_{1}$ and
$\bm{\rho}_{1}$, which results in
\begin{align}
  \sum _{\beta} & \left [ \mathcal{T}_{\alpha \beta}
    + \mathcal{V}_{\alpha \beta} ( E ) 
    + ( M_{\bar{K} \bar{D} N} - E ) \mathcal{N}_{\alpha \beta} \right ]
  \mathcal{C}_{k, n n^{\prime}}^{c} = 0 ,
  \label{eq:three}
\end{align}
where we introduced sets of indices $\alpha = \{ j, \, m, \,
m^{\prime}, \, a \}$ and $\beta = \{ k, \, n, \, n^{\prime}, \, c \}$,
and define $\mathcal{T}_{\alpha \beta}$, $\mathcal{V}_{\alpha \beta}$,
and $\mathcal{N}_{\alpha \beta}$ as
\begin{align}
  \mathcal{T}_{\alpha \beta} \equiv & \delta _{j k}
  \int d^{3} \lambda _{1} \, d^{3} \rho _{1} \,
  \exp \left ( - \frac{\lambda _{a}^{2}}{r_{m}^{2}}
  - \frac{\rho _{a}^{2}}{r_{m^{\prime}}^{2}}
  - \frac{\lambda _{c}^{2}}{r_{n}^{2}}
  - \frac{\rho _{c}^{2}}{r_{n^{\prime}}^{2}} \right )
  \notag \\ & 
  \times \left [ \frac{1}{\mu _{c}^{\prime} r_{n}^{2}}
    \left ( 3 - \frac{2 \lambda _{c}^{2}}{r_{n}^{2}} \right )
    + \frac{1}{\mu _{c} r_{n^{\prime}}^{2}}
    \left ( 3 - \frac{2 \rho _{c}^{2}}{r_{n^{\prime}}^{2}} \right ) \right ] ,
\end{align}
\begin{align}
  \mathcal{V}_{\alpha \beta} ( E )
  \equiv & \int d^{3} \lambda _{1} \, d^{3} \rho _{1} \,
  \exp \left ( - \frac{\lambda _{a}^{2}}{r_{m}^{2}}
  - \frac{\rho _{a}^{2}}{r_{m^{\prime}}^{2}}
  - \frac{\lambda _{c}^{2}}{r_{n}^{2}}
  - \frac{\rho _{c}^{2}}{r_{n^{\prime}}^{2}} \right )
  \notag \\ & \times 
  V_{j k} ( E ) , 
\end{align}
\begin{align}
  \mathcal{N}_{\alpha \beta} \equiv \delta _{j k}
  \int d^{3} \lambda _{1} \, d^{3} \rho _{1} \,
  \exp \left ( - \frac{\lambda _{a}^{2}}{r_{m}^{2}}
  - \frac{\rho _{a}^{2}}{r_{m^{\prime}}^{2}}
  - \frac{\lambda _{c}^{2}}{r_{n}^{2}}
  - \frac{\rho _{c}^{2}}{r_{n^{\prime}}^{2}} \right ) ,
\end{align}
respectively.  We can regard Eq.~\eqref{eq:three} as a generalized
eigenvalue problem of linear algebra.  We numerically solve this to
evaluate the eigenvalue $E = E_{\rm pole}$ and eigenvector
$\mathcal{C}_{j, n n^{\prime}}^{c}$.

\section{\boldmath $\bar{K} \bar{D} N$ molecular state}
\label{sec:4}

\subsection{Eigenenergy}

\begin{figure}[!t]
  \centering
  \PsfigA{8.6cm}{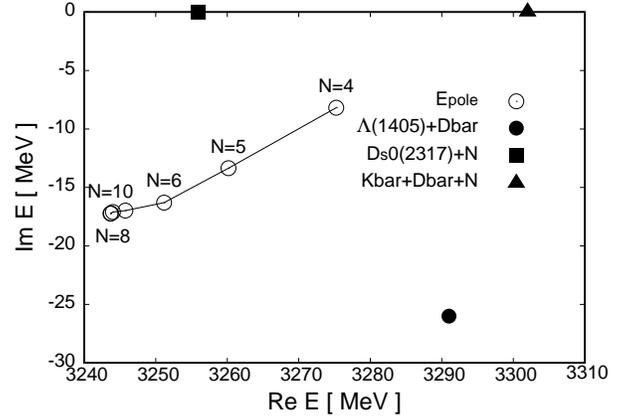}
  \caption{Eigenenergy $E_{\rm pole}$ as a function of the number of
    the Gaussian expansion $N$ (open circles) from $N = 4$ to $N =
    10$.  We also plot the threshold points for the $\Lambda (1405) +
    \bar{D}$, $D_{s 0} (2317) + N$, and $\bar{K} + \bar{D} + N$ as
    filled symbols.}
  \label{fig:trace}
\end{figure}

Now we solve the \Schr equation~\eqref{eq:three} in the $K^{-}
\bar{D}^{0} p$-$\bar{K}^{0} D^{-} p$-$\bar{K}^{0} \bar{D}^{0} n$
coupled channels and search for the $\bar{K} \bar{D} N$ bound state.
For the study of the $\bar{K} \bar{D} N$ system, we fix $r_{\rm min} =
0.1 \fm$ and $r_{\rm max} = 20.0 \fm$.  Taking the number of the
expansion $N = 10$, we find a solution of Eq.~\eqref{eq:three} with
its eigenenergy $E_{\rm pole} = 3244 - 17 i \mev$.  The convergence of
the expansion can be checked by the trace of the eigenenergy $E_{\rm
  pole}$ from $N = 4$ to $10$, which is plotted in the complex energy
plane of Fig.~\ref{fig:trace}.  As one can see, we achieve the
convergence of the expansion with $N \ge 8$.

The real part of the eigenenergy $E_{\rm pole}$ is below the $D_{s 0}
(2317) + N$ threshold ($3256 \mev$) as well as below the $\Lambda
(1405) + \bar{D}$ and $\bar{K} \bar{D} N$ thresholds ($3291 - 26 i
\mev$ and $3302 \mev$, respectively).  This means that this state is
indeed a $\bar{K} \bar{D} N$ quasibound state which cannot decay into
$D_{s 0}(2317) + N$, $\Lambda (1405) + \bar{D}$, nor $\bar{K} \bar{D}
N$.  The binding energy of the quasibound state is $58 \mev$ measured
from the $\bar{K} \bar{D} N$ threshold, $48 \mev$ from the $\Lambda
(1405) \bar{D}$ threshold, and $13 \mev$ from the $D_{s 0} (2317) N$
threshold.

The imaginary part of the eigenenergy indicates decay of the
quasibound state, as we introduced complex-valued potentials
reflecting implicit decay channels such as $\pi \Sigma$ in $\bar{K}
N$.  We emphasize that the imaginary part of the eigenenergy is
obtained in a full calculation rather than in a perturbative one.
From the eigenenergy, we find that the decay width of the quasibound
state is $- 2 \times \text{Im} E_{\rm pole} = 34 \mev$.  The decay of
the quasibound state will be discussed in the next subsection.

\subsection{Decay}

Because $\Lambda (1405)$ decays into $\pi \Sigma$ and $D_{s
  0}(2317)^{-}$ into $\pi \bar{D}_{s}$, we expect that the main decay
channels of the $\bar{K} \bar{D} N$ quasibound state may be $\pi
\Sigma + \bar{D}$ and $\pi \bar{D}_{s} + N$.  To check this, we
perform the same three-body calculations but without the imaginary
parts of the two-body potentials.

When we neglect the imaginary part of the $\bar{K} N$ potential, we
obtain the bound-state eigenenergy at $3240 - 0 i \mev$.  On the other
hand, when we neglect the imaginary part of the $\bar{K} \bar{D}$
($\bar{D} N$) potential, we obtain the bound-state eigenenergy at
$3244 - 18 i \mev$ ($3245 - 20 i \mev$).  These results indicate that
the decay width of the $\bar{K} \bar{D} N$ quasibound state originates
from the $\bar{K} N$ interaction.

Hence, one could conclude that the decay is dominated by the $\pi
\Sigma + \bar{D}$ channel, but it is not the all of the main decay
modes.  As we will see in the next subsection, the $\bar{K} \bar{D} N$
quasibound state has a significant $\bar{K} N ( I = 1 )$ component.
Furthermore, the $\bar{K} N$ effective potential has similar values
of the imaginary parts for the $I = 0$ and $I = 1$ channels (see
Figs.~7 and 8 of Ref.~\cite{Miyahara:2015bya}).  These indicate that
the bound-state decay originating from the $\bar{K} N$ interaction
includes not only the $\pi \Sigma ( I = 0 ) + \bar{D}$ mode but also
the $\pi \Lambda + \bar{D}$ and $\pi \Sigma ( I = 1 ) + \bar{D}$
modes.  As a consequence, the main decay modes of the $\bar{K} \bar{D}
N$ quasibound state are the $\pi \Lambda + \bar{D}$ and $\pi \Sigma +
\bar{D}$ channels.

We here note that the decay width of the $\bar{K} \bar{D} N$
quasibound state, $34 \mev$, is smaller than that of the $\Lambda
(1405)$ as the $\bar{K} N$ quasibound state, $\sim 50 \mev$.  If the
$\bar{K} N$ subsystem in the $\bar{K} \bar{D} N$ quasibound state
behaved like the $\Lambda (1405)$, the decay width of the $\bar{K}
\bar{D} N$ quasibound state would be similar to the width of the
$\Lambda (1405)$.  Indeed, this reduction of the decay width is caused
by the three-body dynamics, in particular the slight extension of the
$\bar{K} N$ distance in the $\bar{K} \bar{D} N$ quasibound state
compared to that in the $\Lambda (1405)$, as we will see in the next
subsection.  From this result we can say that, in general, the decay
width of a three-body quasibound state $A B C$ is not the sum of the
decay widths of the two-body quasibound states $A B$, $B C$, and $C
A$ but it depends on the internal structure of the three-body
quasibound state.

We may consider the two-hadron decay modes, $\bar{K} \bar{D} N \to
\bar{D} \Lambda$, $\bar{D} \Sigma$, and $\bar{D}_{s} N$, as well,
which are not included in our formulation.  For these two-hadron decay
modes, we can use the same argument as in Ref.~\cite{Jido:2008kp} (see
also Fig.~4 therein).  First, the transition to two-hadron states via
a contact interaction is strongly suppressed for a three-body
quasibound state, because the three constituents should meet at a
point for the contact interaction to take place.  Second, the
transition to two-hadron states via virtual meson exchanges is also
suppressed due to the dilute nature of the three-body quasibound
state.  Such a virtual meson exchange process is expected to take
place in the nonmesonic decay of $\bar{K}$-nucleus
systems~\cite{Sekihara:2009yk, Sekihara:2012wj}.  Therefore, in
analogy to the nonmesonic decay of $\bar{K}$-nucleus systems, we can
estimate that the branching ratio of the two-hadron decays of the
$\bar{K} \bar{D} N$ quasibound state will be $\sim 20 \%$, about as
large as the empirical values of the branching ratio of the nonmesonic
decay of $\bar{K}$-nucleus systems.

\subsection{Structure}

Then, we investigate the internal structure of the $\bar{K} \bar{D} N$
quasibound state by using the wave function $\Psi _{j} (
\bm{\lambda}_{1} , \, \bm{\rho}_{1} )$ which is normalized as
\begin{equation}
  \sum _{j = 1}^{3} \int d^{3} \lambda _{1} \, d^{3} \rho _{1} \,
  \left [ \Psi _{j} ( \bm{\lambda}_{1} , \, \bm{\rho}_{1} )
    \right ] ^{2} = 1 .
  \label{eq:norm}
\end{equation}
We emphasize that, because the $\bar{K} \bar{D} N$ quasibound state is
a resonance, we calculate the complex value squared of the wave
function rather than the absolute value squared to normalize the
resonance wave function $\Psi _{j}$ as a Gamow vector.

\begin{table}[!t]
  \caption{Isospin components $X$ and averages of the distances $d$
    for two hadrons in the $\bar{K} \bar{D} N$ quasibound state.}
  \label{tab:comp}
  \begin{ruledtabular}
    \begin{tabular}{lc}
      $X_{\bar{K} N ( I = 0 )}$ & $0.24 + 0.02 i$ 
      \\
      $X_{\bar{K} N ( I = 1 )}$ & $0.76 - 0.02 i$ 
      \\
      $d_{\bar{K} N}$ & $1.13 - 0.39 i \fm$ 
      \\
      \hline
      $X_{\bar{K} \bar{D} ( I = 0 )}$ & $0.98 - 0.01 i$
      \\
      $X_{\bar{K} \bar{D} ( I = 1 )}$ & $0.02 + 0.01 i$
      \\
      $d_{\bar{K} \bar{D}}$ & $0.79 - 0.05 i \fm$
      \\
      \hline
      $X_{\bar{D} N ( I = 0 )}$ & $0.27 - 0.02 i$
      \\
      $X_{\bar{D} N ( I = 1 )}$ & $0.73 + 0.02 i$ 
      \\
      $d_{\bar{D} N}$ & $1.05 - 0.35 i \fm$
    \end{tabular}
  \end{ruledtabular}
\end{table}

We first perform the isospin decomposition.  To this end, we construct
the projection operator to the $\bar{K} N ( I = 0 )$ state as
\begin{equation}
  \mathcal{P}_{\bar{K} N ( I = 0 )} =
  \frac{1}{2} \ket{ K^{-} p + \bar{K}^{0} n }
  \bra{ K^{-} p + \bar{K}^{0} n } .
\end{equation}
By using this projection operator, we can calculate the fraction of
the $\bar{K} N ( I = 0 )$ component in the $\bar{K} \bar{D} N$
quasibound state as
\begin{align}
  X_{\bar{K} N ( I = 0 )} \equiv &
  \braket{\Psi | \mathcal{P}_{\bar{K} N ( I = 0 )} | \Psi }
  = \frac{\left \langle \Psi _{1}^{2} + 2 \Psi _{1} \Psi _{3}
    + \Psi _{3}^{2} \right \rangle}
  {2} ,
\end{align}
where 
\begin{align}
  \left \langle \Psi _{j} \Psi _{k} \right \rangle
  = \int d^{3} \lambda _{1} \, d^{3} \rho _{1} \,
  \Psi _{j} ( \bm{\lambda}_{1} , \, \bm{\rho}_{1} )
  \Psi _{k} ( \bm{\lambda}_{1} , \, \bm{\rho}_{1} ) ,
\end{align}
while the $\bar{K} N ( I = 1 )$ component is 
\begin{equation}
  X_{\bar{K} N ( I = 1 )} = 1 - X_{\bar{K} N ( I = 0 )} .
\end{equation}
Similarly, we can express the projection operators and fractions of
the components for other states as
\begin{equation}
  \mathcal{P}_{\bar{K} \bar{D} ( I = 0 )} =
  \frac{1}{2} \ket{ K^{-} \bar{D}^{0} + \bar{K}^{0} D^{-} }
  \bra{ K^{-} \bar{D}^{0} + \bar{K}^{0} D^{-} } ,
\end{equation}
\begin{align}
  X_{\bar{K} \bar{D} ( I = 0 )} \equiv &
  \braket{\Psi | \mathcal{P}_{\bar{K} \bar{D} ( I = 0 )} | \Psi }
  = \frac{\left \langle \Psi _{1}^{2} + 2 \Psi _{1} \Psi _{2}
    + \Psi _{2}^{2} \right \rangle}{2} ,
\end{align}
\begin{equation}
  X_{\bar{K} \bar{D} ( I = 1 )} = 1 - X_{\bar{K} \bar{D} ( I = 0 )} ,
\end{equation}
\begin{equation}
  \mathcal{P}_{\bar{D} N ( I = 0 )} =
  \frac{1}{2} \ket{ \bar{D}^{0} n - D^{-} p }
  \bra{ \bar{D}^{0} n - D^{-} p } ,
\end{equation}
\begin{align}
  X_{\bar{D} N ( I = 0 )} \equiv &
  \braket{\Psi | \mathcal{P}_{\bar{D} N ( I = 0 )} | \Psi }
  = \frac{\left \langle \Psi _{2}^{2} - 2 \Psi _{2} \Psi _{3}
    + \Psi _{3}^{2} \right \rangle}{2} ,
\end{align}
\begin{equation}
  X_{\bar{D} N ( I = 1 )} = 1 - X_{\bar{D} N ( I = 0 )} ,
\end{equation}
respectively.  The results of the fractions $X$ are listed in
Table~\ref{tab:comp}.  All the fractions are complex because the
quasibound state is a resonance.  Nevertheless, the $\bar{K} \bar{D}$
component in isospin $I = 0$ is very close to unity with small
imaginary part, which implies the dominant $\bar{K} \bar{D} ( I = 0 )$
component inside the $\bar{K} \bar{D} N$ quasibound state.  This is a
consequence of the three-body dynamics to maximize the attraction
among three constituents.  Namely, as one can see from
Fig.~\ref{fig:Ueff}, the $\bar{K} \bar{D} ( I = 0 )$ interaction is
most attractive among the pairs of the two constituents in the present
formulation, and $\bar{K} N ( I = 0 )$ comes next.  Besides, in
contrast to the moderate $\bar{K} N ( I = 1 )$ attraction, the
$\bar{K} \bar{D} ( I = 1 )$ interaction is negligible, as the $\bar{K}
\bar{D} (I = 1)$ interacts only through the $\pi \bar{D}_{s}$
intermediate channel [see Eq.~\eqref{eq:V_KDKD1}].  Therefore, the
three-body dynamics increase the $\bar{K} \bar{D} ( I = 0 )$ fraction
as much as possible so as to maximize the attraction in the $\bar{K}
\bar{D} N$ quasibound state.  In this sense, the $\bar{K} \bar{D}$
interaction, both the $I = 0$ and $1$ components, is most essential
for the internal structure of the $\bar{K} \bar{D} N$ quasibound
state.  Furthermore, we have checked that the following relations
hold:
\begin{align}
  \left \langle \Psi _{1}^{2} \right \rangle \approx
  \left \langle \Psi _{1} \Psi _{2} \right \rangle \approx
  \left \langle \Psi _{2}^{2} \right \rangle \approx \frac{1}{2} , 
\end{align}
\begin{align}
  \left \langle \Psi _{1} \Psi _{3} \right \rangle \approx
  \left \langle \Psi _{2} \Psi _{3} \right \rangle \approx
  \left \langle \Psi _{3}^{2} \right \rangle \approx 0 .
\end{align}
These relations indicate that the quasibound state is indeed described
by the $[ \bar{K} \bar{D} (I = 0) ] p$ configuration.  The negligible
contribution from the $\bar{K}^{0} \bar{D}^{0} n$ channel also
explains the results that the $\bar{K} N ( I = 0 )$ and $\bar{D} N ( I
= 0 )$ components are close to $1/4$.

\begin{figure}[!t]
  \centering
  \PsfigA{8.6cm}{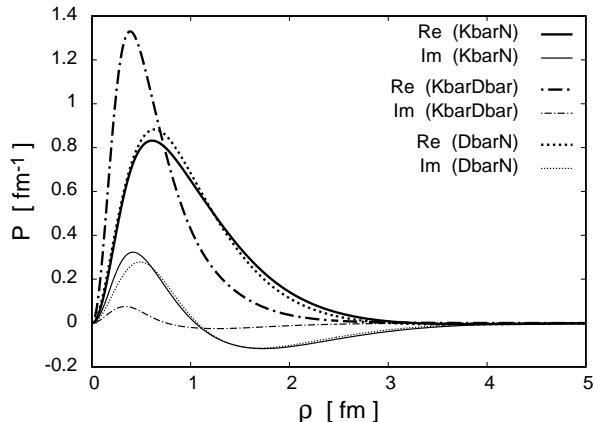}
  \caption{Density distributions for the $\bar{K} N$, $\bar{K}
    \bar{D}$, and $\bar{D} N$ subsystems in the $\bar{K} \bar{D} N$
    quasibound state.}
  \label{fig:density}
\end{figure}

Next we investigate how the two-hadron subsystems behave in the
$\bar{K} \bar{D} N$ quasibound state by calculating the density
distribution for each pair of two constituents, which is defined as
\begin{align}
  & \text{P}_{\bar{K} N} ( \rho _{2} )
  \equiv \rho _{2}^{2} \int d \Omega _{\bm{\rho} _{2}} \, d^{3} \lambda _{2}
  \, \sum _{j = 1}^{3} 
  \left [ \Psi _{j} ( \bm{\lambda}_{1} , \, \bm{\rho}_{1} ) \right ] ^{2} ,
  \\
  & \text{P}_{\bar{K} \bar{D}} ( \rho _{3} )
  \equiv \rho _{3}^{2} \int d \Omega _{\bm{\rho} _{3}} \, d^{3} \lambda _{3}
  \, \sum _{j = 1}^{3} 
  \left [ \Psi _{j} ( \bm{\lambda}_{1} , \, \bm{\rho}_{1} ) \right ] ^{2} ,
  \\
  & \text{P}_{\bar{D} N} ( \rho _{1} )
  \equiv \rho _{1}^{2} \int d \Omega _{\bm{\rho} _{1}} \, d^{3} \lambda _{1}
  \, \sum _{j = 1}^{3} 
  \left [ \Psi _{j} ( \bm{\lambda}_{1} , \, \bm{\rho}_{1} ) \right ] ^{2} ,
\end{align}
where $\Omega _{\bm{\rho}}$ is the solid angle of the vector
$\bm{\rho}$.  The integration of $\text{P}_{\bar{K} N}$,
$\text{P}_{\bar{K} \bar{D}}$, and $\text{P}_{\bar{D} N}$, with respect
to $\rho _{2}$, $\rho _{3}$, and $\rho _{1}$, respectively, in the
range $[ 0, \, \infty )$ is unity according to the
  normalization~\eqref{eq:norm}.\footnote{The measures of the Jacobi
    coordinates satisfy $d^{3} \lambda _{1} \, d^{3} \rho _{1} = d^{3}
    \lambda _{2} \, d^{3} \rho _{2} = d^{3} \lambda _{3} \, d^{3} \rho
    _{3}$.}  The resulting density distributions for the pairs of two
  constituents are plotted in Fig.~\ref{fig:density}.  From the
  figure, the $\bar{K} N$ and $\bar{D} N$ distributions are similar to
  each other and extend typical hadronic scale $1 \fm$.  However, the
  $\bar{K} \bar{D}$ distribution in the $\bar{K} \bar{D} N$ quasibound
  state is significant only below $1 \fm$, which is similar to the
  $\bar{K} \bar{D}$ distribution in the $D_{s 0}(2317)$ in
  Fig.~\ref{fig:KD-DN}.  This result supports the $[ \bar{K} \bar{D}
    (I = 0) ] p$ configuration for the $\bar{K} \bar{D} N$ quasibound
  state.  Furthermore, because the $\bar{K} \bar{D} ( I = 0 )$
  subsystem is compact, the distributions of the $\bar{K} N$ and
  $\bar{D} N$ in the $\bar{K} \bar{D} N$ quasibound state are similar
  to each other.

From the density distributions, we can calculate the averages of the
distances between two constituents:
\begin{align}
  & d_{\bar{K} N} \equiv 
  \sqrt{\int _{0}^{\infty} d \rho _{2} \, \rho _{2}^{2} 
    \text{P}_{\bar{K} N} ( \rho _{2} ) } ,
  \\ &
  d_{\bar{K} \bar{D}} \equiv 
  \sqrt{\int _{0}^{\infty} d \rho _{3} \, \rho _{3}^{2} 
    \text{P}_{\bar{K} \bar{D}} ( \rho _{3} )} ,
  \\
  & d_{\bar{D} N} \equiv 
  \sqrt{\int _{0}^{\infty} d \rho _{1} \, \rho _{1}^{2} 
    \text{P}_{\bar{D} N} ( \rho _{1} )} .
\end{align}
The results are listed in Table~\ref{tab:comp}.  As one can see, the
averages of the distances have small imaginary parts due to the
resonance nature but their real parts are dominant.  Therefore, below
we focus on the real parts of the distances.  Among the three
distances, the distance between $\bar{K} \bar{D}$ is the smallest,
which indicates the compact $\bar{K} \bar{D} ( I = 0 )$ subsystem.
The distance between $\bar{K} \bar{D}$ in the $\bar{K} \bar{D} N$
quasibound state is smaller than that in the $D_{s 0}(2317)$ as the
$\bar{K} \bar{D}$ two-body bound state, $0.93 \fm$.  This is because
the $N$ assists the $\bar{K} \bar{D}$ attraction in the $\bar{K}
\bar{D} N$ quasibound state via the $\bar{K} N$ and $\bar{D} N$
interactions.  The distance between $\bar{D} N$ in the $\bar{K}
\bar{D} N$ quasibound state becomes much smaller than that of the
$\bar{D} N$ two-body bound state, $3.66 \fm$, owing to the compact
$\bar{K} \bar{D}$ subsystem and the strong attraction between $\bar{K}
N$.  We also note that, although the distance between $\bar{K} N$ in
the $\bar{K} \bar{D} N$ quasibound state is similar to that in the
$\Lambda (1405)$, the $\Lambda (1405)$ is not effective degrees of
freedom in the $\bar{K} \bar{D} N$ quasibound state because the
isospin component $X_{\bar{K} N ( I = 0 )}$ is only about $1/4$.

In terms of the structure, we can understand the decrease of the decay
width of the $\bar{K} \bar{D} N$ quasibound state compared to the
$\Lambda (1405)$.  This is caused by the fact that the $\bar{K} N$
distance in the $\bar{K} \bar{D} N$ quasibound state is slightly
larger than that in the $\Lambda (1405)$, which is crucial to the
decay width of the $\bar{K} \bar{D} N$ quasibound state originating
from the $\bar{K} N$ interaction.  Actually, because the $\bar{K} N$
effective potential has a finite range, which is $b = 0.38 \fm$ in our
study, the increase of the $\bar{K} N$ distance directly reduces the
probability of overlapping $\bar{K} N$ for the decay.  As a
consequence, we obtain the smaller decay width of the $\bar{K} \bar{D}
N$ quasibound state than that of the $\Lambda (1405)$.  We note that
the isospin structure of the $\bar{K} \bar{D} N$ quasibound state,
i.e., dominant $\bar{K} N ( I = 1 )$ component, is irrelevant to the
decrease of the decay width, because the $\bar{K} N$ effective
potential takes similar values of the imaginary parts for the $I = 0$
and $I = 1$ channels.

\subsection{Theoretical ambiguities}

Finally, we discuss theoretical ambiguities in our scenario of the
generation of the $\bar{K} \bar{D} N$ quasibound state.

As we have seen, the $\bar{K} \bar{D} N$ quasibound state is generated
by the two kinds of strong attraction, the $\bar{K} N ( I = 0 )$
interaction and $\bar{K} \bar{D} ( I = 0 )$ interaction, and moderate
$\bar{D} N$ attraction.  Among them, the $\bar{D} N$ interaction is
not well determined due to poor experimental data.  To check the
influence of the $\bar{D} N$ interaction, we switch off the $\bar{D}
N$ interaction and perform the three-body calculations.  As a result,
we obtain the $\bar{K} \bar{D} N$ quasibound state with eigenenergy
$3248 - 21 i \mev$, whose value is similar to the full-calculation
value $3244 - 17 i \mev$.  Therefore, we can say that ambiguity of
the $\bar{D} N$ interaction is irrelevant.

Besides, although the $\bar{K} \bar{D}$ interaction is fixed to
reproduce the $D_{s 0}(2317)$ as the $\bar{K} \bar{D}$ bound state, it
is not clear how much the $D_{s 0}(2317)$ contains a ``bare'' $s
\bar{c}$ component rather than the $\bar{K} \bar{D}$ molecular
component.  Such an $s \bar{c}$ component will weaken the attraction
of the $\bar{K} \bar{D}$ effective potential and may affect our
scenario.  However, the $D_{s 0}(2317)$ generated in the present
formulation already contains some missing-channel contribution rather
than the $\bar{K} \bar{D}$-$\eta \bar{D}_{s}$ channels via the
intrinsic energy dependence of the interaction~\eqref{eq:V_KDKD0}.
Actually, the intrinsic energy dependence of the $\bar{K}
\bar{D}$-$\eta \bar{D}_{s}$ interaction introduces missing-channel
fraction $= 22 \%$ to the $D_{s 0}(2317)$, as seen in
Ref.~\cite{Navarra:2015iea}, which can be interpreted as a bare $s
\bar{c}$ component.  Therefore, our scenario allows the $D_{s
  0}(2317)$ to have $\sim 20 \%$ fraction of the bare $s \bar{c}$
component.

Our scenario would be affected by the treatment of the energy
dependence of the two-body interaction in the three-body dynamics [see
  Eqs.~\eqref{eq:eps_DN}, \eqref{eq:eps_KN}, and \eqref{eq:eps_KD}].
To check this, we firstly fix two-body energy of only one of the three
pairs in the three-body system to its threshold energy while we keep
the energy dependence for other two pairs.  When we keep the energy
dependence for the $\bar{K} \bar{D}$ and $\bar{D} N$ potentials but
fix the $\bar{K} N$ energy as $\epsilon _{\bar{K} N} = m_{K} + m_{N}$,
we obtain the eigenenergy $3242 - 21 i \mev$.  Similarly, when we fix
the $\bar{K} \bar{D}$ ($\bar{D} N$) energy as $\epsilon _{\bar{K}
  \bar{D}} = m_{K} + m_{D}$ ($\epsilon _{\bar{D} N} = m_{D} + m_{N}$),
we obtain the eigenenergy $3238 - 18 i \mev$ ($ 3231 - 25 i \mev$).
The shifts of the eigenenergy in these cases are not significant
because the energy dependence of the two-body effective potentials is
not essential in the energy region of interest, i.e., around their
thresholds.  Secondly, if we fix all the two-body energies to the
two-body threshold energies, $\epsilon _{\bar{D} N} = m_{D} + m_{N}$,
$\epsilon _{\bar{K} N} = m_{K} + m_{N}$, and $\epsilon _{\bar{K}
  \bar{D}} = m_{K} + m_{D}$, the eigenenergy becomes $3219 - 34 i
\mev$.  Thirdly, when we fix the two-body energies to the pole
positions of the two-body bound states, i.e., $\epsilon _{\bar{K} N} =
1424 - 26 i \mev$, $\epsilon _{\bar{K} \bar{D}} = 2317 \mev$, and
$\epsilon _{\bar{D} N} = 2805 \mev$, the eigenenergy becomes $3226 -
28 i \mev$.  These treatments bring more biding energy and width to
the $\bar{K} \bar{D} N$ state, but the $\bar{K} \bar{D} N$ quasibound
state exists in any case.

From the above discussions, we conclude that the $\bar{K} \bar{D} N$
quasibound state will exist even if we take into account theoretical
ambiguities.

\section{Summary and concluding remarks}
\label{sec:5}

In this study we investigated the $\bar{K} \bar{D} N$ quasibound state
with spin/parity $J^{P} = 1/2^{+}$ and isospin $I = 1/2$ in a
nonrelativistic three-body potential model.  Following the approach in
Refs.~\cite{Hyodo:2007jq, Miyahara:2015bya} for the $\bar{K} N$
effective local potential, we constructed the $\bar{K} \bar{D}$ and
$\bar{D} N$ effective local potentials based on phenomenological
models, with which we obtained the $D_{s 0}(2317)$ as a $\bar{K}
\bar{D}$ bound state and $\bar{D} N$ bound state, respectively.  These
two-body effective potentials implicitly contain inelastic channels.
In particular, the inclusion of the open channels is essential to
describe the $\bar{K} \bar{D} N$ system as a decaying state.

By solving the three-body \Schr equation in the $K^{-} \bar{D}^{0}
p$-$\bar{K}^{0} D^{-} p$-$\bar{K}^{0} \bar{D}^{0} n$ coupled channels
with the constructed two-body effective local potentials, we obtained
the $\bar{K} \bar{D} N$ quasibound state with eigenenergy $3244 - 17 i
\mev$.  The real part of the eigenenergy is below the $D_{s 0} (2317)
+ N$ and $\Lambda (1405) + \bar{D}$ thresholds as well as the $\bar{K}
\bar{D} N$ threshold, so the $\bar{K} \bar{D} N$ quasibound state
cannot decay into $D_{s 0}(2317) + N$, $\Lambda (1405) + \bar{D}$, nor
$\bar{K} \bar{D} N$.  From the imaginary part of the eigenenergy, we
calculated the decay width of the $\bar{K} \bar{D} N$ quasibound state
into $\pi \bar{D} \Lambda$, $\pi \bar{D} \Sigma$ to be $34 \mev$.  In
addition to the three-hadron decay modes, the $\bar{K} \bar{D} N$
quasibound state will have two-hadron decay modes $\bar{K} \bar{D} N
\to \bar{D} \Lambda$, $\bar{D} \Sigma$, and $\bar{D}_{s} N$, and the
branching ratio of the two-hadron decays was estimated to be $\sim 20
\%$.

As for the internal structure, the $\bar{K} \bar{D} N$ quasibound
state takes $[ \bar{K} \bar{D} (I = 0) ] p$ configuration with compact
$\bar{K} \bar{D}$ subsystem because this configuration can maximize
the attraction among three constituents by utilizing the strong
$\bar{K} \bar{D} ( I = 0 )$ attraction fully.  We found that the
three-body dynamics may increase distance between two constituents and
hence may reduce decay width of the three-body quasibound state
compared to that of the two-body quasibound state of the constituents,
as is the relation between the $\bar{K} \bar{D} N$ quasibound state
and $\bar{K} N$ quasibound state [$\Lambda (1405)$].  We also
discussed theoretical ambiguities, and conclude that the $\bar{K}
\bar{D} N$ quasibound state will exist even if we take into account
theoretical ambiguities.

Finally, we remark the possibility of the experimental search for the
$\bar{K} \bar{D} N$ quasibound state.  Because the $\bar{K} \bar{D} N$
quasibound state has both strangeness $S = -1$ and charm $C = -1$,
practical candidate is the production in relativistic heavy-ion
collisions~\cite{Cho:2010db, Cho:2011ew, Cho:2017dcy}.  One can search
for the $\bar{K} \bar{D} N$ quasibound state in, e.g., the $\pi
\bar{D} \Lambda$ and/or $\bar{D} \Lambda$ invariant-mass spectra of
relativistic heavy-ion collisions.  The $B$ meson decays in $B$
factories are feasible as well.  In this case, for instance, the
$\bar{B}_{s} ( s \bar{b} ) \to \pi \bar{D} \Lambda + \bar{p}$,
$\bar{D} \Lambda + \bar{p}$ processes are suitable.

\begin{acknowledgments}

  The authors acknowledge E.~Hiyama for fruitful discussions on the
  few-body calculations.  The authors are grateful to D.~Jido for
  helpful discussions on dynamics which emerges uniquely in three-body
  systems.
  This work was partly supported by JSPS KAKENHI Grants No.~JP15K17649
  and No.~JP18K13545.
  
\end{acknowledgments}

\end{document}